# Examining the correlation of the level of wage inequality with labor market institutions

*By* **Virginia TSOUKATOU** †

**Abstract.** Technological change is responsible for major changes in the labor market. One of the offspring of technological change is the SBTC, which is for many economists the leading cause of the increasing wage inequality. However, despite that the technological change affected similarly the majority of the developed countries, nevertheless, the level of the increase of wage inequality wasn't similar. Following the predictions of the SBTC theory, the different levels of inequality could be due to varying degrees of skill inequality between economies, possibly caused by variations in the number of skilled workers available. However, recent research shows that the difference mentioned above can explain a small percentage of the difference between countries. Therefore, most of the resulting inequality could be due to the different ways in which the higher level of skills is valued in each labor market. The position advocated in this article is that technological change is largely given for all countries without much scope to reverse. Therefore, in order to illustrate the changes in the structure of wage distribution that cause wage inequality, we need to understand how technology affects labor market institutions.In this sense, the pay inequality caused by technological progress is not a phenomenon we passively accept. On the contrary, recognizing that the structure and the way labor market institutions function is largely influenced by the way institutions respond to technological change, we can understand and maybe reverse this underlying wage inequality. Consequently, we would like to examine to what extent the reason behind the increase of wage inequality in some countries but not in others is attributed to the structure and the way the institutions of labor market work. In this article, we will attempt to examine this hypothesis by empirically elaboratingon the relationship between SBTC, inequality and labor market institutions.

## 1. Introduction

The causes of rising wage inequality in recent decades have been the subject of research in the area of labor economics. Technological change has brought about major changes in the labor market. The result of technological change is the SBTC where for many researchers is the cause of rising wage inequality (Krueger, 1993). Although technology has affected all developed countries to a similar degree, the magnitude of the increase in wage inequality was not similar. Following the predictions of the SBTC theory, the different levels of inequality could be due to varying degrees of skill inequality between economies, possibly caused by variations in the number of skilled workers available. However, recent research (Blum & Guerin-Pace, 2000; Devroye & Freeman, 2001) shows that the difference mentioned above can explain only 7% of the difference between different countries. Therefore, most of the resulting inequality could be due to the different ways in which the higher level of skills are valued in each labor market.

† Department of Political Science and International Relation, University of Peloponnese, Greece.
☎. +248 370-2125 ✉. vtsoukatou@uop.gr

In this prism, the research hypothesis we will consider is that the reason that wage inequality has not increased in all industrialized countries in the same way is that the diversity of labor market institutions is responsible for the different valuation of skills in the economy, by squeezing distribution of earnings in the labor market and thus making technological change less skill-biased. Technological progress in this sense is more endogenous and the assumption that it is solely responsible for pay inequality has a fatalistic connotation. The position advocated in this article is that technological change is largely given for all countries without much scope to reverse. Therefore, in order to illustrate the changes in the structure of wage distribution that cause wage inequality, we need to understand how technology affects labor market institutions.

In this sense, the pay inequality caused by technological progress is not a phenomenon we passively accept. On the contrary, recognizing that the structure and the way labor market institutions function is largely influenced by the way institutions respond to technological change, we can understand and maybe reverse this underlyingwage inequality.

In the following sections, we will attempt to examine the aforementioned research hypothesis by examining the relationship between SBTC and inequality and the relationship of labor market institutions (minimum wage, degree of centralization of wage bargaining and the role of wage labor) with the level of pay inequality. The choice of these institutions for consideration is not random. The fact that there is a correlation between minimum wage and inequality is widely accepted in the literature. We, therefore, want to examine this correlation empirically as the existence of a minimum wage sets a binding lower pay level that affects the level of wage inequality for low-wage (and usually low-skilled) workers.

The degree of centralization of wage bargaining likewise appears from the literature to be largely related to the reduction in the extent of wage distribution causing a reduction in inequality. Finally, the decline in membership of labor unions and the reduction of their influence on wage setting, particularly during the 1980s, has been linked to increased wage inequality, particularly in the US and the United Kingdom (Card, 1992; DiNardo, Fortin, & Lemieux, 1996; Fortin & Lemieux, 1997; Freeman, 1991; Lee, 1999). Of course, one could also mention other labor market institutions that may affect wage inequality, but the ones mentioned above are the most prevalent.In conclusion, we will empirically investigate all these labor market institutions by examining their correlation with the level of inequality in eight OECD countries in order to identify the institutional causes behind the increase in inequality observed in recent decades.

## 2. Theory and literature

### 2.1. SBTC as a cause of the increasing wage inequality

As noted above, inequality can be either the result of technological change that is causing changes in the demand for skills in the labor market or due to institutional factors. According to the first reason, wage inequality is the way the labor market reacts to new working conditions caused by changes in the skills of workers. The SBTC is the result of a growing demand for highly skilled workers at the expense of the demand for unskilled labor. In this way, companies are forced to expand the payroll of their employees in order to attract the specialized employees they need.

In this new wage distribution created, some workers (usually the most skilled) benefit because their marginal productivity is higher than others. This high inequality is difficult to be reduced without losing part of the efficiency of the labor market, as trying to reduce it would require reforms that would increase unemployment and affect competitiveness due to higher wage costs affecting unit costs of the product. This is due to the fact that countries that allow higher wage inequality at the lower level of wage distribution (usually translated into lower wages) usually achieve higher levels of employment.Until the 1980s, wage inequality remained largely flat. As Blinder observed (1980, p.2) the inequality in the United States in 1977 was almost similar to that in 1947. However, the increase in the levels of pay inequality became apparent in the 1980s. This rise was initially not easily understood from where it came from as some researchers even considered it a consequence of the deep recession of 1981-1982.

The causes behind rising inequality began to crystallize in the 1990s, where a series of research articles linking rising inequality to the SBTC was released (Autor, Katz, & Krueger, 1997; Bound & Johnson, 1992; Juhn, Murphy, & Pierce, 1993; Katz & Murphy, 1992; Krueger, 1993; Levy & Murnane, 1992).

According to the SBTC theory, by using technology can produce more product with the same number of production factors. This increase in efficiency is what drives the technological progress of skilled labor, which greatly increases the demand for workers with specialized knowledge and skills and thus increases their earnings by widening the inequality of employees' labor income.

The theoretical starting point of this literature began with the empirical observation that the supply of skilled labor and its remuneration increase simultaneously only if the demand for skilled labor increases significantly. This increase in demand is a factor that demonstrates the beginnings of SBTC. All of the above combined with the time when the rise of inequality began (a few years after the creation of personal computers in the early 1980s) gave rise to the prevailing view that the emergence of new technologies caused the increase in the demand for skilled labor, magnifying the pay gap. This theory continued to monopolize scientific debate until the early 1990s.

However, looking at the issue with a time gap of almost three decades since the 1990s when scientific research was first published on the relationship of inequality with the SBTC, we note that this view does not explain as one would expect the increasing inequality over the years followed. Although technological breakthroughs have continued rapidly since the 1990s (in many cases the 1990s were accompanied by more significant technological improvements than the 1980s. A typical example is the "Internet revolution" characterized by the growing use of the Internet), but the pattern of wage inequality has changed, since the level inequality has stabilized in most non-Anglo-Saxon countries. There has been a clear differentiation between developed economies since the 1980s. Although countries such as France, Germany or Japan have experienced similar technological transformations with the United States, there has not been a similar increase in inequality.

Given that technological change is something that has affected all developed countries, we would expect inequality to increase in every country. However, as we can see in this was not the case.

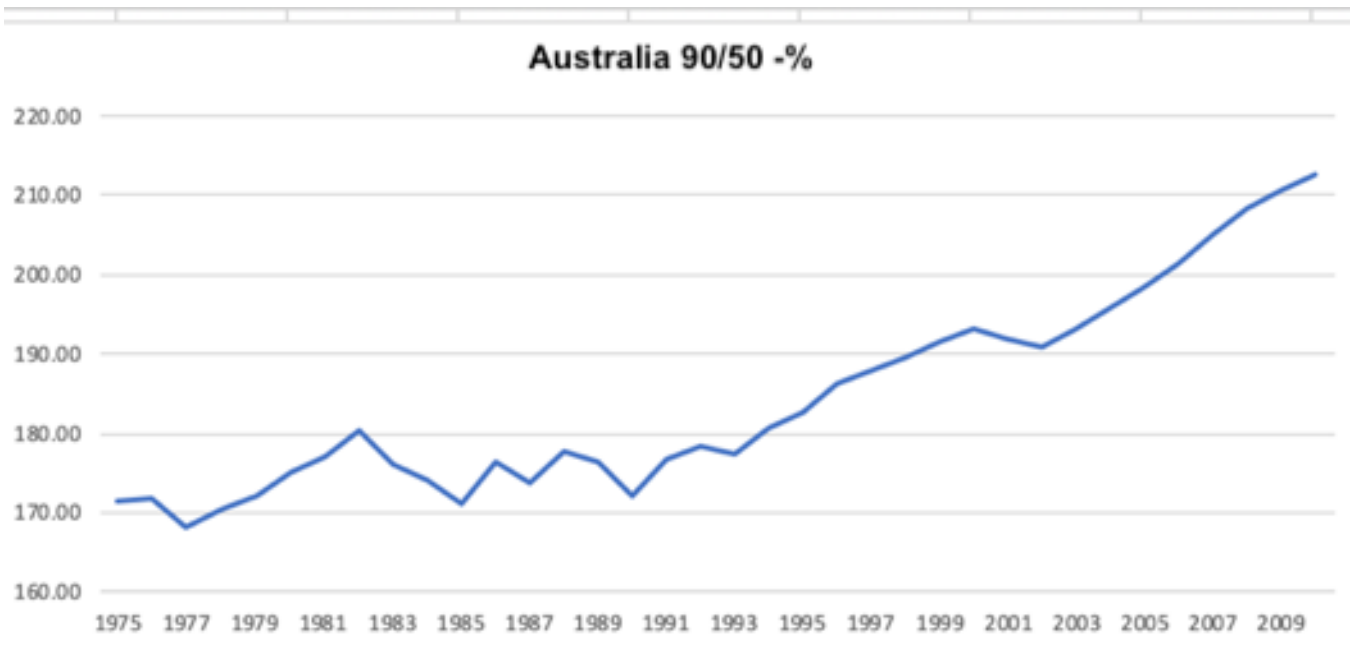

**Figure 1.**

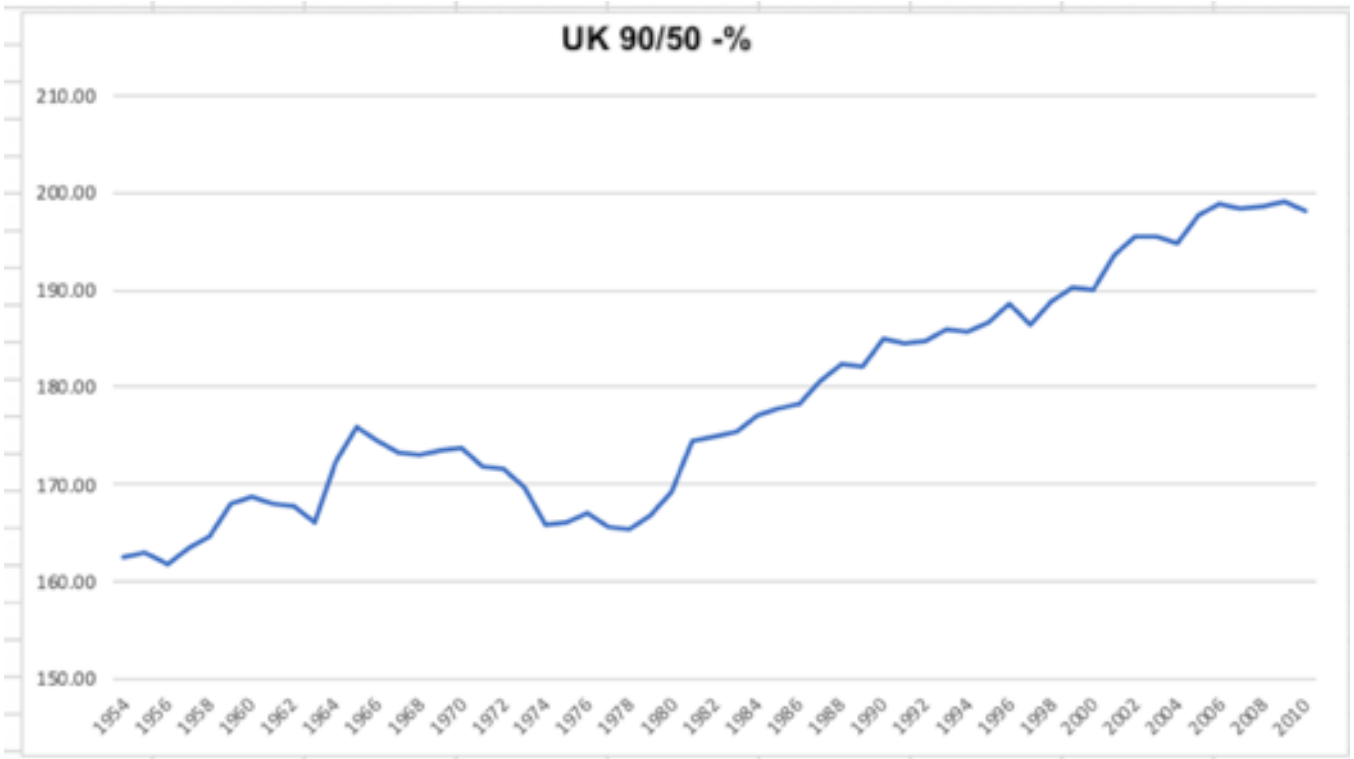

**Figure 2.**

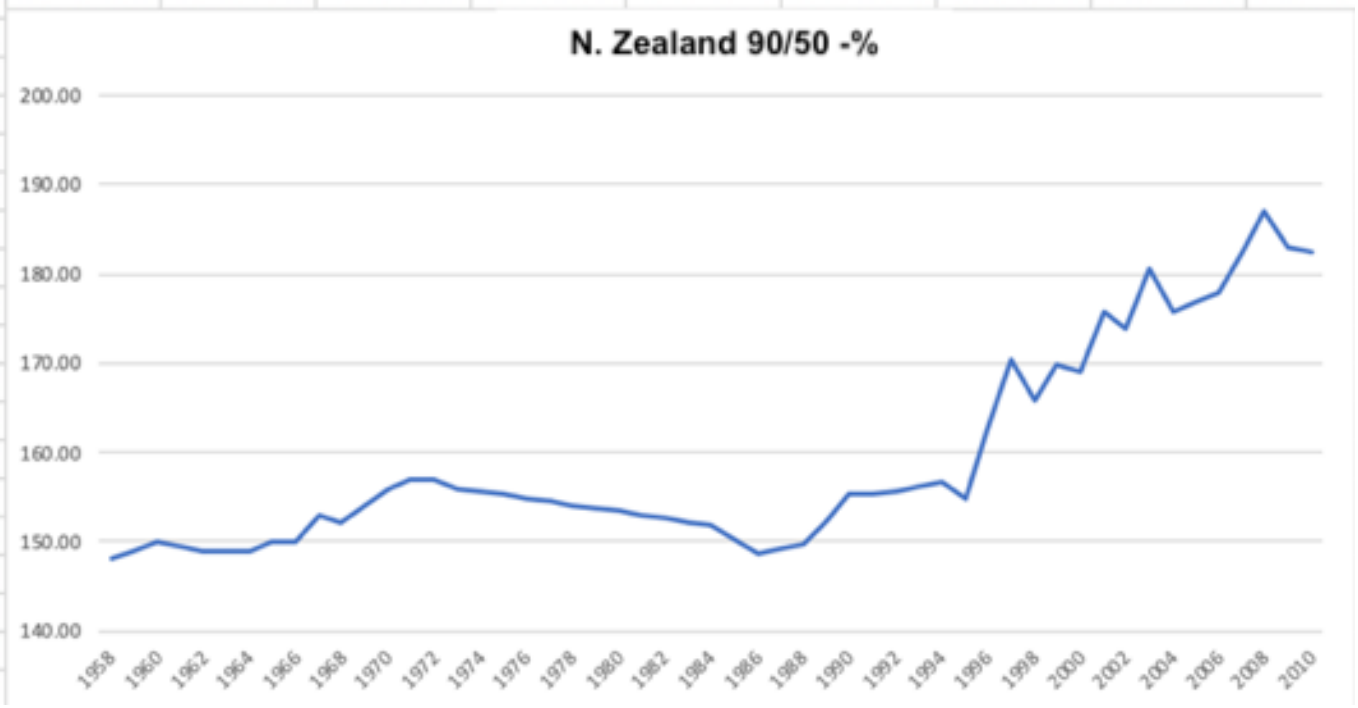

Figure 3.

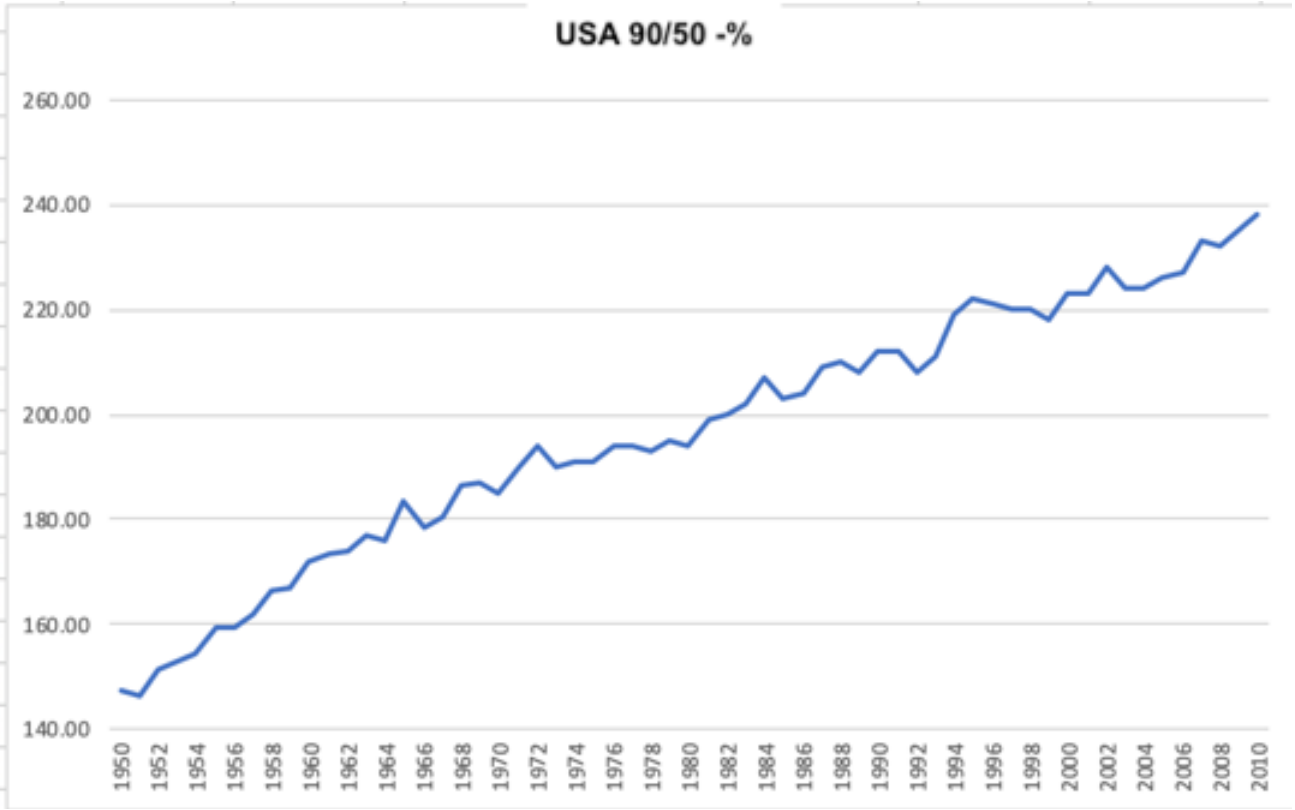

Figure 4.

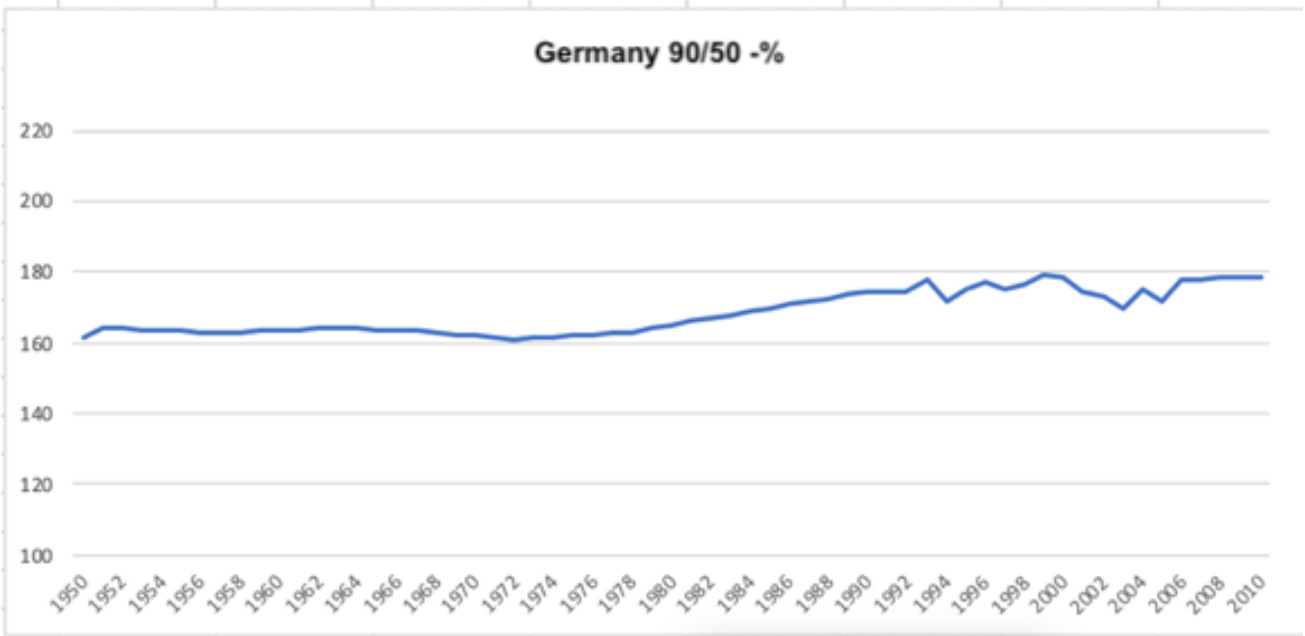

Figure 5.

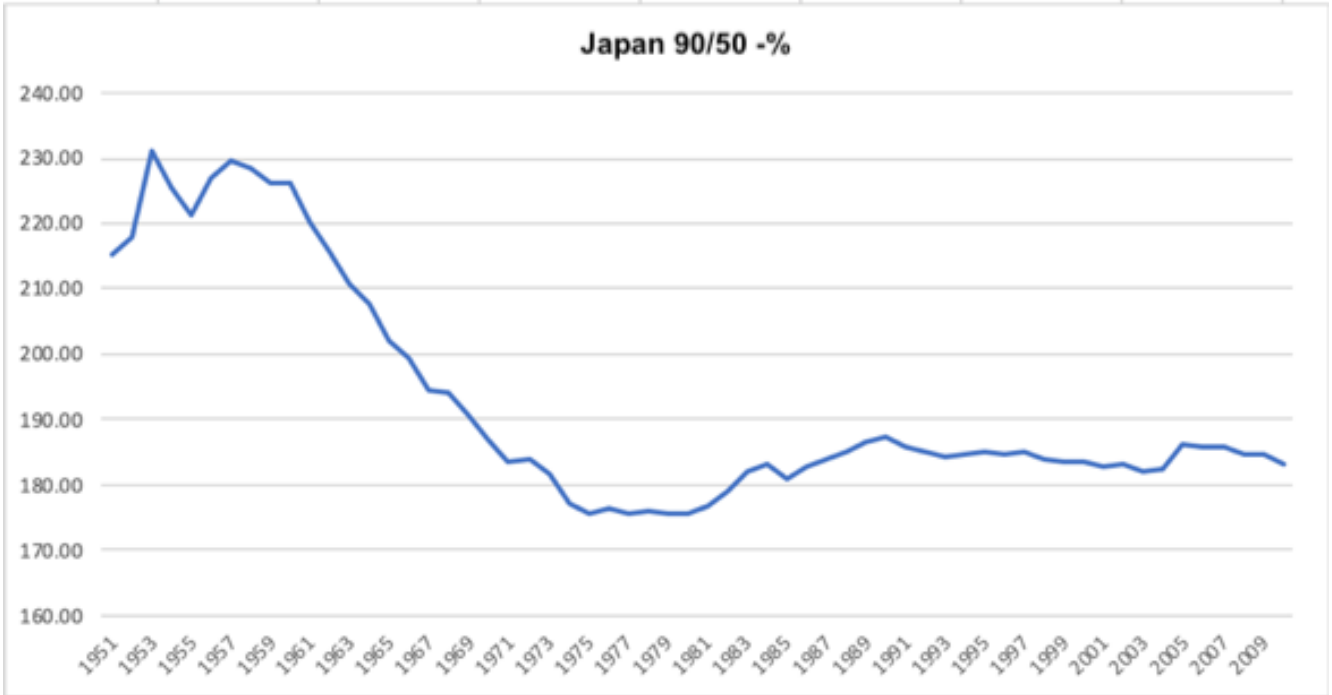

Figure 6.

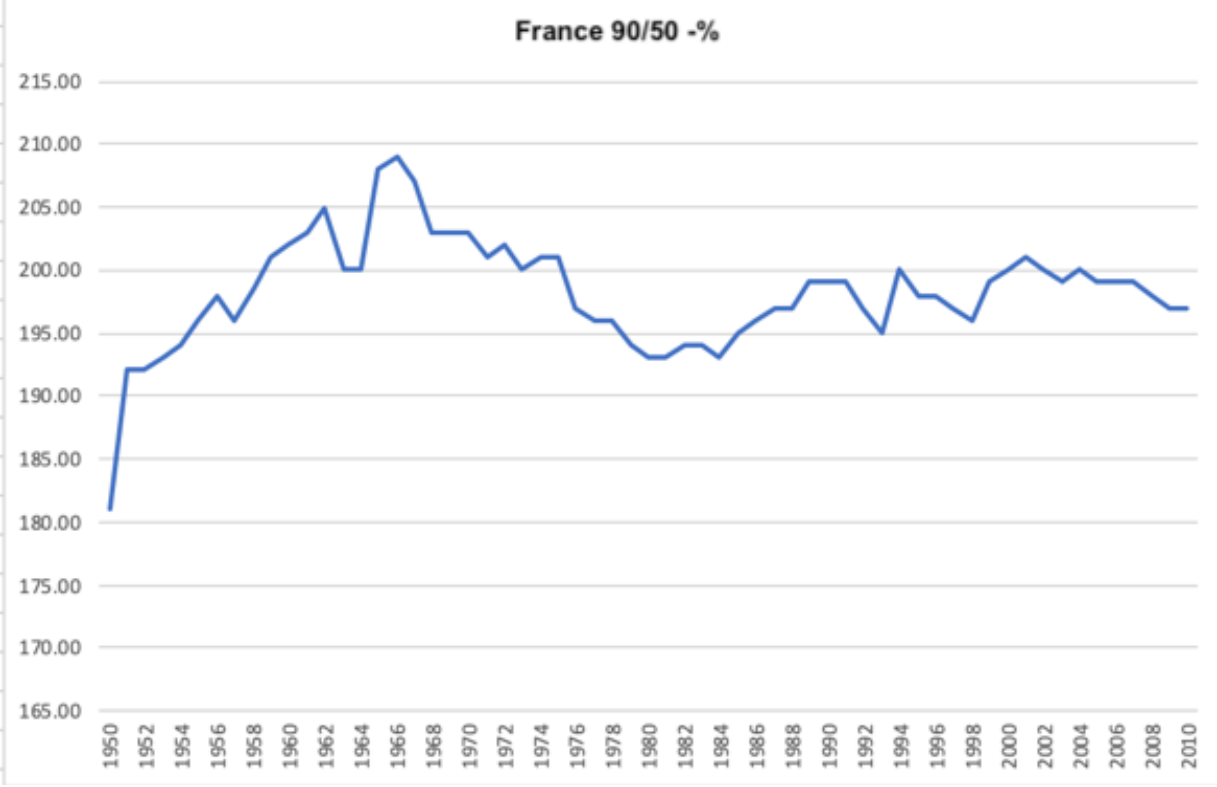

**Figure 7.**

In Figures 1-7 we have shown how much higher is the percentage of a worker earning 90% of the payroll than someone earning 50% in 1950-2010 (the year of the start of the chart varies according to available per country data). The data source is the database of Atkinson *et al.*, (2017). In the years when no data were available, we estimated prices assuming an equal annual change over the interim years. The reason chosen for examining this point of pay distribution is to better illustrate the impact of technological change on the pay gap in different countries.For this reason, we use as an indicator how much higher the remuneration of a worker who is at 90% of the payroll than a worker at 50%, because technological change differentiates the level of demand for high skills and thus how they are paid. As a rule of thumb, skilled workers are in the middle and above the payroll. Therefore, looking at the highest end of the distribution we can graphically illustrate how SBTC relates to inequality in countries with inherent differences in their labor market structure. To illustrate the labor hypothesis that increasing inequality is not primarily a result of the SBTC per se, but relates to how the labor market responds to technological change due to the diversity of its institutions, we look at the examples of seven developed OECD countries (Australia, New Zealand, USA, France, Germany, Japan and the United Kingdom). All of these countries were developed, so technological change greatly affected them in the same way.

However, by looking at the graphs we see that inequality has not increased in the same way in all cases. Looking at the charts of Australia, the US, the United Kingdom and New Zealand we see a sharp increase in high income inequality in line with what we would expect in the case of wage inequality caused by the SBTC.In contrast, in other countries (Germany, France, Japan) the level of wage inequality remains largely stable, with no linear increase in inequality as a result of technological change. From these graphs, we can see that in the first group of countries (Australia, New Zealand, USA and the United Kingdom) higher skilled workers were receiving higher and higher salaries compared to middle-skilled workers. In the second group of countries (Germany, France and Japan) we find that the skills of skilled workers were not compensated as disproportionately high as those of middle skilled workers, demonstrating that technological change either does not cause SBTC and consequently increases inequality in each case, or there are other factors in specific countries that reduce the SBTC level.As mentioned above, the SBTC has similarly affected all technologically advanced countries, but the effect on the level of inequality was not similar. From one perspective one could argue that this may be due to the fact that while technological change has created a greater demand for skilled labor, the number of skilled workers that the education system could provide to the economy varied on a case by case basis.

As a result, there has been a significant shift in the demand for skills that have favored more skilled workers. In countries where the level of remuneration was more flexible (such as in the US), the remuneration of the less skilled fell sharply as a result of the reduced demand for low-skilled workers. In countries where the structure of labor market institutions was more robust, wage inequality did not increase to such an extent. However, in many cases, unemployment of the less skilled increased (Nickell & Bell, 1996). Of course one could argue that this is not the case in cases like Germany or the Netherlands for example where although the minimum wage is comparatively higher than the US the unemployment is not significantly different. This could be explained by the different way the education system operates, which creates specific minimum acceptable skill levels for each student, resulting in a situation that even the less skilled workers having a relatively high level of productivity compared to countries such as the US or the United Kingdom (Marx, 2007). In this context, in economies where the number of more skilled workers was small, their pay would be higher and the higher wages would increase inequality (Katz & Murphy, 1992; Murphy, Riddell, & Romer, 2003).

Newer research (Blum & Guerin-Pace, 2000; Devroye & Freeman, 2001), however, shows that this difference in the ability of different systems to produce skilled workers can account for only 7% of the pay gap between countries. Correspondingly, 36% of this inequality for the US can be explained by a higher skill premium paid to more skilled workers compared to the less skilled.One logical question therefore, is why countries value skills and human capital so differently. Our theory is that the diversity of labor market institutions squeezes the pay range in a way that makes technological change less skill-based. Technological progress in this sense has a more endogenous character. Correspondingly, the assumption that it is primarily responsible for wage inequality leaves the impression that inequality is timeless and little can be done to reverse it. On the contrary, the causes of divergent pay disparity should be sought in the other institutions of the labor market and in their ability to respond to a variety of negative shocks. Anglo-Saxon countries such as the US, Australia, New Zealand and the United Kingdom are characterized by decentralized wage bargaining (and hence no or low minimum wage) and powerless labor unions, among others. Consequently, a negative impact on the demand for unskilled labor (which could be caused by technological change, for example) would have the effect of lowering the salaries of less skilled workers. Countries with more centralized wage bargaining, higher statutory minimum wage and stronger labor unions would react differently to a negative demand shock (mainly of the less skilled workers) because their wage levels would remain unchanged, keeping the wage inequality stable.

Even in the United States, it appears that the degree of participation in labor unions is largely related to inequality. According to Card (1992), the decline in participation causes a 20% increase in men's salary fluctuations in 1980. Besides, the reduction in the real minimum wage is important in increasing disparity since most of the inequality is at the lower end of the wage distribution in the US (Atkinson, Hasell, Morelli, & Roser, 2017; Card & DiNardo, 2002; Lemieux, 2008). Looking at the above as a whole, we find that the explanation of the SBTC as the main cause of the inequality observed between 1980 and 1990 is less convincing than the diversity of labor market institutions. In the following sections, we will examine the labor market institutions separately in order to try to assess the impact of each on increasing inequality.

### 2.2. Institutional factors as a cause of wage inequality

The second broader category of causes behind the growing inequality are labor market institutions. As it was clear from the previous section, SBTC's argument as the main cause of pay inequality ceases to be particularly convincing. Countries with a more concentrated wage distribution usually also have a more concentrated distribution of skills in the labor market. At first glance we would assume that this observation is compatible with the SBTC but looking deeper into the problem we would observe that perhaps the most important 'crack' in this view is the case of the Scandinavian countries. In the US and the rest of the Anglo-Saxon countries, both wage distribution and skills fluctuations have been observed at the same time as a result of the better return on investment of skills of workers. However, in the Scandinavian countries, although the level of skills is particularly broad, the level of wage inequality is one of the lowest in the world (therefore the wage distribution is quite compressed). This may be related to the nature of the education system operating differently than in the Anglo-Saxon countries.In addition, how workers' skills are valued is different due to differences in the structure of labor market institutions in countries with low inequality. In this light, institutions such as the minimum wage, the role of trade unions and the structure of wage bargaining should be considered. Although we try to look at the SBTC separately from the labor market institutions, the interaction between them and the overall skill level is undeniable.

Labor market institutions can influence the distribution of skills in the economy when, for example, higher tax rates are raised. So if it were consolidated it could change the decisions of the younger generations about the level of skills they want to acquire. This is because educational decisions are largely made with the available employment opportunities and the extent to which pay increases with education. Thus, even in this way, the direction that technological change can followmay change (Barany, 2011; Broecke, Quintini, & Vandeweyer, 2015).

### 2.3. Minimum wage and wage inequality

The level of the minimum wage can affect the level of pay inequality in a variety of ways, either negatively or positively. For example:

• It may affect the direction of technological change as mentioned above by intervening in the way low-skilled workers are paid. This will reduce the supply of skilled labor by condensing wage distribution and making technological change less skill-based. Of course, such a decision would have a negative impact on other factors in the economy, making it potentially less competitive and efficient.

• When the level of the minimum wage is extremely low, its impact on reducing wage inequality will be very limited. However, we should keep in mind that a reasonable increase in the minimum wage could reduce inequality. However, this applies up to a certain level of minimum wage because if it exceeds the maximum that the economy can afford, inequality will increase due to the unemployment that will be caused.

The two basic theories that explain the importance of minimum wage inequality are the redistributive theory (Freeman, 1996) and the theory of marginal productivity.According to redistributive theory, the level of the minimum wage can affect inequality at the lower end of the wage distribution in three ways:

The first is through consumer products produced by workers who are paid the minimum wage. An increase in the minimum wage will cause an increase in the cost of production and ultimately will influence the price of the product produced. Consequently, the increase in the purchasing power of low-wage workers comes through a reduction in the purchasing power of other citizens, affecting the level of overall inequality in the wage distribution. The second way that the minimum wage may affect wage inequality according to the redistributive theory is by reducing the income of shareholders of businesses employing low-wage workers. This could be the case if the increased production costs due to the increase in the minimum wage that was previously passed on to product buyers were borne by the companies. In this way, by lowering the profitability of the companies, the shareholders or executives' incomes are reduced, thus indirectly reducing inequality between high and low pay levels. Lastly, the minimum wage affects wage inequality through unemployment that may cause its increase. A reasonable increase in the minimum wage could improve prosperity but when the increase exceeds the maximum tolerance levels for the economy then the result will be an increase in unemployment and a consequent increase in inequality (Litwin, 2015).

The second theory that attempts to capture the relationship between minimum wage and inequality is the theory of marginal productivity. According to this theory, the pay gap between employees is caused by their different skills. Skilled employees or otherwise more qualified receive a higher salary. The theory we are examining recognizes three different levels of competence from which three scenarios emerge: low, moderate and high skills. In this model of economy that we are considering as an example, there may be a "covered" and "uncovered" sector in the sense that the "uncovered" may either have fewer or no benefits in the event of unemployment than the "covered" (e.g. fixed-term workers who may not be entitled to unemployment benefit, self-employed or other categories who for any reason are paid below the minimum wage, etc.) (Litwin, 2015). In this economy that we are examining as an example to understand the theory of marginal productivity, the workers who are paid the minimum wage but who work in the "covered sector" are those with moderate skills who are most affected by the fluctuation of the minimum wage. The lowest-skilled workers are absorbed into the "uncovered sector". If the minimum wage increases, then the demand for work in the covered sector' increases on the part of the workers, thereby reducing the demand for the uncovered. As a result increasing the wages there. This increase in the salaries of non-skilled workers reduces inequality without decreasing the salaries of skilled workers. But if the increase in the minimum wage in the "covered sector" is greater than the economy can afford, it will cause unemployment, causing many workers to seek work in the "uncovered sector", reducing the wages there. In this case, the disparity between low-skilled and highly skilled workers will widen. Finally, if the covered sector' increases substantially the minimum wage and creates unemployment but the wage in the uncovered sector' is clearly lower than the wage demand of the new unemployed coming from the covered sector then due to reduced labor supply they will increase the salaries of workers in the "uncovered sector" reducing inequality (Litwin, 2015).

Since a significant number of OECD countries whether they have statutory minimum wages or the majority workforce is covered by collective agreements, it is not very common for a large proportion of workers to be employed in an "uncovered sector". In this sense, the number of workers who are paid below the minimum wage is not large enough to affect the inequality in the ways mentioned. Therefore, the redistributive theory seems to prevail.

Based on the aforementioned points, we see that the minimum wage should, in theory, be related to the level of wage inequality in the economy. The level of the minimum wage is largely influenced by how the wage is set. The diversity in the ways in which minimum wages are determined from country to country explains the existence of so many different trajectories of wage inequality. In summary, the statutory minimum wage is more likely to reduce the range of wage distribution and thus wage inequality due to its legally binding nature which makes it less likely to be circumvented than a minimum wage determined by collective bargaining. Because the minimum wage is by definition the same for all workers in the economy, irrespective of the sector in which they are employed or their jobs, there is less cross-sectoral wage inequality.

When the minimum wage is determined by collective bargaining with binding nature for all workers in the industry, then the levels of wage inequality are comparable (but possibly slightly worse) to those that would result through the prism of a legally set minimum wage. The problem, however, is that in many cases not all workers in the industry are bound by the individual agreement signed by the labor union. In this case, inequality will increase because workers who are not members of the labor union will be paid lower than those who are members.

### 2.4. Correlation between the level of wage bargaining and the level of wage inequality

Another factor affecting the wage distribution is the degree of centralization of payroll negotiations. To a large extent, the increasing inequality observed in recent years is due to the decentralized trends in most countries worldwide. To make the labor market more flexible and reduce unemployment rates, governments are trying to decentralize wage bargaining to the detriment of the general level of wage inequality.According to Calmfors & Driffill's (1988) hypothesis, the degree of centralization of wage bargaining is an important determinant of a country's macroeconomic performance and competitiveness in the sense that it affects the level of real wages and thus the unemployment rate. According to Calmfors and Driffill's theory, centralized and decentralized wage bargaining performs better than sectoral wage bargaining, in terms of competitiveness and unemployment. The theoretical starting point of Calmfors and Driffill's research was the observation that countries such as Sweden, Austria or Norway had very good employment performance. Their explanation for this phenomenon was based on the existence of a centralized wage bargaining system. At the same time, in countries with decentralized wage bargaining (for example, the United Kingdom or Canada) even if they had higher unemploymentthe labor market flexibility was superior.

These observations led the authors to conclude that the two ends (either decentralization at the operational level or centralization at the national level) work better than intermediate forms of negotiation (sectoral negotiations). The explanation given for this observation is that in the case of centralization, there are large and centralized labor unions that understand their particular strengths and take into account the impact of their wage effects on unemployment and inflation.

On the other hand, in decentralized bargaining systems where bargaining takes place within each business, labor unions have very limited power that does not allow them to have excessive demands. In the case of wage bargaining at the sectoral level, labor unions have considerable power but operating in an environment larger than that of a business but smaller than that of a national economy leads them to ignore the macroeconomic consequences of their claims, believing that they do not affect the rest of the economy. In this light, Calmfors and Driffill have argued that on the one hand centralization acts as a way of taking into account the broader interests, while on the other hand decentralization necessarily forces actors to limit their requirements to follow the peculiarities of the market. In both cases, however, wage restraint is achieved in both cases. Observing the graphical representation of this finding in a graph, we will see that it is hump-shaped.

According to Calmfors & Driffill's (1988) theory, a decentralized and centralized wage bargaining system could have the same effects on employment. However, in their theory, they did not mention the factor of wage inequality. These two payroll systems have different wage distributions, with the result that the overall level of wage inequality in the economy also differs.

When wage bargaining is centralized, wage inequality tends to be lower for two reasons: the first is ideological and the second is political. The ideological starts from the theoretical starting point that in collective bargaining, workers' representatives are interested in having as much justice as possible in the way members of the labor union are paid. This in itself does not alter the level of inequality in payroll systems as it would potentially apply even when there is total decentralization. The point is, however, that looking at this issue in terms of human resources management would have a significant impact on profitability, productivity and in general the attitude of employees towards the business. In a centralized wage bargaining institutional environment, the sense of fairness that results from the structure of wage distribution has an impact not only on the staff of an individual business but on a large portion of the workforce as a whole.

Consequently, this sense of justice is central during a centralized negotiation resulting in a more concentrated pay distribution (Wallerstein, 1999). This preference of labor unions for condensing skills pay and productivity is partly the reason that there is a lower degree of inequality in high incomes than in decentralized payroll systems. An extension thus, this explains the increase in high income inequality in countries with weak labor unions (Lemieux, 2008).

The second factor that holds back inequality when wage negotiations are centralized is political. A centralized wage bargain involving numerous smaller labor unions differentiates each one's influence on wage setting. Consequently, precisely because the leaders of each labor union wish to be re-elected, the average voter in the wage setting is taken into account. The result of this move is to increase the remuneration of average voters by increasing their remuneration to the detriment of high salaries. This attracts those below the middle of the wage distribution as well, resulting in reduced pay inequality (Wallerstein, 1999).

To a large extent, and because of Europe's economic crisis, in recent years countries have been trying to reform their wage bargaining systems in order to reduce unemployment. In general, the more decentralized systems take fuller account of the financial conditions prevailing in each business. This determines the wage of employees based on the real potential of the economy, industry or business respectively. Precisely because the particularities of each business are taken into account (either positive such as the increased productivity a company may have, or negative) in determining the amount of wage unemployment tends to be much lower. The price for this, however, is the widening of the pay gap and the consequent increase in inequality.

The effects of technological change on inequality have begun to be seen since 1980. In Germany, an increase in high income inequality became apparent in late 1980and later in 1990, it began to expand at a slower pace than the US and at the bottom of the wage distribution. This shift in inequality and the widening of wage distribution coincides with the institutional reforms that have begun in the German labor market. Until that time, they kept inequality at a lower level and differentiated in the way the wage distribution would respond in comparison to the Anglo-Saxon countries. However, the price for this lower wage inequality was higher unemployment. In order to increase the flexibility of the labor market, Germany has introduced reforms aimed at entering the labor market of the economically inactive population (which were largely paid with low salaries resulting in increasing inequality). At the same time, the power of trade unions was reduced and the remuneration became clearly connected with the personal productivity resulting in increased inequality at the top ends of the wage distribution (Dell Aringa & Pagani, 2005).

### 2.5. The relation between trade unions and wage inequality

By 1970, the prevailing view was that labor unions were increasing inequality. This, as we shall see below, has a solid theoretical basis but is not empirically confirmed. Freeman (1980) has shown that although the action of trade unions has a dual nature in terms of the impact on inequality, their role as a factor in reducing wage inequality prevails. Observing the trajectory of wage inequality in the United States and the United Kingdom, we can see that these two countries have seen a huge increase in inequality at the same time as the decline in the number of workers being members oftrade unions.

The role of trade unions in wage inequality can have a dual nature: either negatively or positively. On the one hand, labor unions increase the earnings of their members more than the earnings of non-members by exerting pressure on employers' representatives. This increase in the remuneration of union members over other workers changes the way remuneration is distributed throughout the economy. When most members of a labor union are highly skilled employees then pay inequality tends to increase. However, in cases where members of the labor union are low-wage workers, wage inequality is reduced (Freeman, 1980). Given the technological change of recent years, it is more likely that members of labor unions tend to be paid salaries at the bottom end of the payroll, negatively affecting inequality. On the other hand, the existence of trade unions can promote wage equality in three ways: by imposing specific pay scales on an independent basis for all employees so that all employees of a company working in the same job function are paid the same. The second way of narrowing the range of wage distribution is by reducing wage differentials between manual and non-manual employees.In many cases, most of the members of the trade unions are manual workers, so the levels of remuneration for non-manual employees are negotiated in such a way that most of the salary increase goes to the most numerous category of workers within the union: that is, the manual workerswith salaries near the bottom of the wage distribution (Asher & DeFina, 1995).

Finally, the third and perhaps most important way to reduce inequality is the imposition by the labor unions of specific levels of remuneration for each job so that all employees covered by the agreement are paid the same as their counterparts in another company if they offer the same services. The establishment of a common wage policy per job directly affects the wage distribution as wage dispersal within the companies involved is reduced. The agreement on uniform pay is a move that is in the interest of both companies and employees, and its applicability remains high. From the business point of view, it is clear that no company would want the remuneration paid to its employees to be higher than the remuneration paid by its competitors for the same job. Similarly, it is pleasurable for workers because they feel a greater degree of pay equity since their salaries are not determined by any other personal characteristics other than their job position. In addition, when there is a common wage cost in an industry, it is more difficult to reduce earnings in order to reduce the competitive price of a product. This may create other problems of flexibility and sustainability but nonetheless, it is a question of the unions being achieved, so their members have no reason not to welcome it.Although both theoretical approaches to labor unions and inequality appear to be convincing, several empirical studies show a reduction in wage inequality when labor unions exist. This is also the reason why the decline in participation rates has been largely linked to the increasing wage inequality observed in many countries in recent decades. Especially in the United States, the decline in employee participation in labor unions is particularly pronounced. Indicatively, it is reported that between 1970 and 1992, employee participation decreased by 50% (from 26% to 13%) (Asher & DeFina, 1995). In general, when an economy is to a small extent exposed to technological change (largely due to its productive structures) then the benefits of an employee from joining trade unions outweigh the limitations of the wage distribution imposed. But when the impact of the SBTC is severe, skilled workers no longer wish to join trade unions. This is because the productivity gap between skilled workers and non-skilled workers is increasing rapidly due to new production technologies making skilled workers more sought after. In this sense, since they know that their work future is secured either by the existing employer or someone else, they do not want to join labor unions that squeeze their earnings for the benefit of their less skilled members (Gordon, 2001). Furthermore, given the existence of technological change in developed countries, we would conclude that most members of labor unions are middle- or low-skilled. Following the reasoning outlined above, precisely because union members are less specialized, their representatives will push for fixing wage levels that reduce inequality and condense wage distribution.

### 2.6. Causes behind the decline of trade union participation rates

Bibliographically, one of the main institutional factors that have contributed to the growing wage inequality has been the decline in labor unions participation rates. As mentioned above, SBTC is an important factor in reducing participation. Skilled employees who know that through their expanded skills are secured their work future do not want to become union members as they limit the range of their remuneration. On the contrary, if they negotiate directly at the company level their remuneration,they are able to agree on significantly higher wages. Another factor contributing to the decline in participation has to do with the structural changes that are gradually transforming economies. In previous years the majority of jobs were in industry / manufacturing where it was easier for workers to join labor unions. On the contrary, in recent years the institutional transition of several economies, from production economies to service economies, has been completed. This is because globalization has drived the production process to less costly developing economies where workers' wages are lower. The result is that the main occupation of most of the workers in these countries has shifted to sectors focusing on providing services where it is more difficult to harmonize the policies imposed by a labor union. Precisely for this reason, participation declines as their narrative is significantly narrower.Finally, another reason for the decline in labor unions is that some of the services previously provided by labor unions are now secured by the state. For example, in some countries in the past, labor unions were tasked with creating security and worker protection rules, distributing unemployment benefits, retirement plans and setting working hours. Consequently, many of the workers who previously joined labor unions for the aforementioned reasons have no reason to continue to be unionized.

## 3. Data set and method

### 3.1. The model

In this section, we present the model we have constructed to test whether the institutional factors examined, in theory, affect wage inequality in practice. The period we are looking at in this empirical study is from 1980 to 2016, given that the 1980s is the decade when the SBTC peaked and there was a significant increase in wage inequality. In order to proceed with the creation of the final multivariate linear regression model to test the influence of each of these variables (union density, degree of centralization, GDP, (deflated) minimum wage and collective bargaining coverage) we used as wage inequality metric the ratios 90/10, 50/10 and 90/50. Besides the labor market institutions, we examined in the previous section, in this model we also use as a variable the level of deflated GDP and the collective bargaining coverage. The deflated GDP was used to have an index that reflects the general state of economic activity. According to Deaton (2013, p.4), economic growth per se since the Industrial Revolution has been associated with increasing inequality. The sample used to create the model includes eight OECD countries (Australia, Belgium, France, USA, Japan, the Netherlands, New Zealand, and Canada) that were industrialized during the period we are examining - so affected largely similar to technological change, they had a statutory minimum wage in place and their labor market institutions were diverse (varying degrees of centralization, union density, and coverage of collective bargaining).

In order to examine the influence of each of these variables as stated above, three multiple linear regressions were applied to substantially check the relationship of one dependent continuous variable each time (the 90/10, 50/10 or 90/50 ratios respectively) and five continuous independent variables (deflated minimum wage, collective bargaining coverage, degree of centralization of wage bargaining, union density and deflated GDP). In our analysis, it was observed that two of the model assumptions (the assumption of regularity and the assumption of homoscedasticity in particular) are violated. More precisely, the Kolmogorov-Smirnov and Shapiro-Wilks tests gave us a p-value of 0.000 <0.05, thus we reject the null hypothesis that the residuals of the model follow a normal distribution. Similarly, in the case of homoscedasticity, the residuals plots appear to have a pattern and not to be random. In order to correct the hypotheses, some transformations were used in either the dependent variable or the explanatory variables, or both types of variables. After testing our model we found that it is best to transform some of the explanatory variables (deflated minimum wage and deflated GDP). As a transformation we used logarithm.

### 3.2. Results

According to the multivariate analysis applied, it was found that all factors had a statistically significant effect on the 90/10 inequality. Specifically, from Table 1 we observe that multiple regression was found to be statistically significant (F (5,277) = 259,055, p-value <0.001), with $R^2$ = 0.821). Regarding the deflated minimum wage, it is observed that one percentage point increase is associated with a reduction of inequality of 90/10 by 0.006 (-0.606*ln($\frac{101}{100}$ units while maintaining the other variables constant.

**Table 1.** *Presenting the results for 90/10*

| Table 1 90/10 | Correlation coefficient | Std. error | t | p-value | 95% confidence Interval Lower bound | Upper bound | VIF |
|---|---|---|---|---|---|---|---|
| Deflated minimum wage, base year 1980_log | -0.606 | 0.022 | -27.128 | **<0.001** | -0.65 | -0.562 | 4.775 |
| Collective bargaining coverage | -0.003 | 0.001 | -2.378 | **0.018** | -0.005 | 0.000 | 4.671 |
| Degree of centralisation of wage bargaining | -0.191 | 0.032 | -6.051 | **<0.001** | -0.254 | -0.129 | 5.742 |
| Union density | 0.011 | 0.002 | 5.981 | **<0.001** | 0.007 | 0.015 | 2.117 |
| Deflated GDP, base year 1980_log | 0.369 | 0.016 | 23.055 | **<0.001** | 0.338 | 0.401 | 5.877 |
| F | 259.055 | | | | | | |
| $R^2$ | 0.824 | | | | | | |
| Adj. $R^2$ | 0.821 | | | | | | |

Regarding collective bargaining coverage, it is observed that for one unit of increase we expect a decrease of inequality of 90/10 by 0.003 units while maintaining the other variables constant. Concerning the degree of centralization of wage bargaining, it is observed that for one unit of increase we expect a reduction of the inequality of 90/10 by 0.191 units keeping the other variables constant. On the contrary, for the union density, it is observed that for one unit of increase we expect an increase of inequality of 90/10 by 0.011 units keeping the remaining variables constant. Finally, we observe that a percentage increase in deflated GDP is expected to increase inequality 90/10 by 0.003 ($0.369*\ln(\frac{101}{100})$) units while maintaining the other variables constant.

**Table 2.** *Presenting the results for 50/10*

| Table 2 50/10 | Correlation coefficient | Std. error | t | p-value | 95% confidence Interval Lower bound | Upper bound | VIF |
|---|---|---|---|---|---|---|---|
| Deflated minimum wage, base year 1980_log | -0.177 | 0.01 | -18.003 | **<0.001** | -0.197 | -0.158 | 4.775 |
| Collective bargaining coverage | -0.001 | 0.001 | -1.75 | 0.081 | -0.002 | 0.000 | 4.671 |
| Degree of centralisation of wage bargaining | -0.111 | 0.014 | -7.931 | **<0.001** | -0.138 | -0.083 | 5.742 |
| Union density | 0.007 | 0.001 | 8.527 | **<0.001** | 0.005 | 0.008 | 2.117 |
| Deflated GDP, base year 1980_log | 0.099 | 0.007 | 13.971 | **<0.001** | 0.085 | 0.112 | 5.877 |
| F | 145.867 | | | | | | |
| $R^2$ | 0.724 | | | | | | |
| Adj. $R^2$ | 0.719 | | | | | | |

According to the multivariate analysis applied, it was found that all factors had a statistically significant effect on the 50/10 inequality. Specifically, from Table 2 we observe that multiple regression was found to be statistically significant ($F(5,277) = 145.867$, p-value <0.001), with $R^2$ = 0.719. Regarding the deflated minimum wage, it is observed that a 1% increase is associated with a reduction of inequality of 50/10 by 0.002 ($-0.177*\ln(\frac{101}{100})$) units keeping the other variables constant. Concerning collective bargaining coverage, it is observed that for one unit of increase we expect a 50/10 inequality to decrease by 0.001 units while keeping the other variables constant.

Regarding the degree of centralization of wage bargaining, it is observed that for one unit of growth we expect a 50/10 inequality to decrease by 0.111 units while keeping the other variables constant. On the contrary, for the union density, it is observed that for one unit of increase we expect an increase of inequality of 50/10 by 0.007 units keeping the other variables constant. According to the literature, there is a correlation between a decrease in union membership and an increase in inequality. Prior empirical research does not normally take into account the years after the late 1990s. This factor may also have influenced the result obtained as a regression coefficient. According to our research, 1 unit of increase in inequality is related to 0.011 units of increase in union density. One hypothesis that could explain this effect of linear regression would be the diversification of labor composition in recent years. In recent years, the general level of education in developed economies has increased, resulting in an increase in the number of skilled workers. As the number of high-skilled employees increases, it is more likely this increase to alter the constitution of the trade unions by having more highly skilled members. In this way, the profile of the 'average worker/member' is also changed and consequently, their role in reducing the extent of pay distribution becomes questionable. Finally, it is observed that for an increase of 1% of deflated GDP we expect an increase of inequality of 50/10 by 0.001 ($0.099*\ln(\frac{101}{100})$) units keeping the other variables constant.

**Table 3.** *Presenting the results for 90/10*

| Table 3 50/10 | Correlation coefficient | Std. error | t | p-value | 95% confidence Interval Lower bound | Upper bound | VIF |
|---|---|---|---|---|---|---|---|
| **Deflated minimum wage, base year 1980_log** | -0.137 | 0.007 | -20.048 | **<0.001** | -0.151 | -0.124 | 4.775 |
| **Collective bargaining coverage** | -0.001 | 0.000 | -1.774 | 0.077 | -0.001 | 0.000 | 4.671 |
| **Degree of centralisation of wage bargaining** | 0.002 | 0.010 | 0.235 | 0.814 | -0.017 | 0.021 | 5.742 |
| **Union density** | -0.004 | 0.001 | -6.668 | **<0.001** | -0.005 | -0.003 | 2.117 |
| **Deflated GDP, base year 1980_log** | 0.092 | 0.005 | 18.831 | **<0.001** | 0.083 | 0.102 | 5.877 |
| **F** | 199.452 | | | | | | |
| **$R^2$** | 0.783 | | | | | | |
| **Adj. $R^2$** | 0.779 | | | | | | |

Finally, in the latter model, according to the multivariate analysis applied, it was found that all factors again have a statistically significant effect on 90/50 inequality. Specifically, from Table 3 we observe that multiple regression was found to be statistically significant ($F (5,277) = 199.452$, p-value <0.001), with $R^2 = 0.783$.

Regarding the deflated minimum wage, it is observed that a 1% increase is associated with a reduction of the 90/50 inequality by 0.001 ($-0.137*\ln(\frac{101}{100})$) units while keeping the other variables constant. Concerning collective bargaining coverage, it is observed that for one unit of increase, the inequality of 90/50 is expected to decrease by 0.001 units while maintaining the other variables constant. On the contrary, regarding the degree of centralization of wage bargaining, it is observed that for one unit of growth we expect an increase of 90/50 inequality by 0.002 units while maintaining the other variables constant. Concerning the union density, it is observed that for one unit of increase, a 90/50 inequality is expected to decrease by 0.004 units while maintaining the other variables constant. Finally, it is observed that a percentage increase in deflated GDP is expected to increase the 90/50 inequality by 0.001 ($0.092*\ln(\frac{101}{100})$) units.

## 4. Findings

This chapter deals with the issue of increasing wage inequality. Since the early 1980s, there has been an increasing wage inequality initially attributed to the impact of technological change and the resulting technological progress of the Skill Biased Technical Change (SBTC). Our working hypothesis was based on the fact that if all industrialized countries were exposed to technological change, the pay gap would have to be affected correspondingly.

This was not the case, and in order to explain this phenomenon, many different causes have been proposed, such as the inherent differences between the educational systems of each country how each economy is valued the highest level of education and skills. Therefore, the hypothesis we examined was that wage inequality depends primarily on how labor market institutions (minimum wage, degree of centralization of wage bargaining, and the role of labor unions) respond to the given technological change and influence how the skills and human capital are valued in general. According to our empirical research, the labor market institutions examined justify 82.1% of the increase in wage inequality (as defined by the wage inequality 90/10) while the remaining 17.9% could be due to other reasons such as the different magnitude of technological change among countries. In different parts of the wage distribution (50/10 and 90/50) the degree of inequality due to labor market institutions is 71.9% and 77.9% respectively. All in all, the main factor in the labor market related to wage inequality is the level of the lower wage (-0.606 in the wage inequality 90/10) followed by (in two of the three models) GDP growth rate (0.369 in 90/10) demonstrating the general level of economic growth. Similarly, the degree of centralization of wage bargaining (-0.191 in the 90/10 distribution) and the trade union density (0.011 in the 90/10 distribution) is a notable factor of increasing inequality in both of the three models. Finally, the least significant factor regarding the level of wage inequality is the degree of coverage of collective bargaining (-0.003 in the wage inequality model 90/10).